\documentclass[prb,eqsecnum,showpacs,preprint]{revtex4}
\usepackage{epsfig,amssymb,amsmath}
\font\bba=msbm10 scaled 1080
\font\bbb=msbm8 
\font\bbc=msbm6 
\newfam\bbfam
\textfont\bbfam=\bba
\scriptfont\bbfam=\bbb
\scriptscriptfont\bbfam=\bbc
\def\bb{\fam\bbfam\bba}
\def\R{{\bb R}}

\begin{document}
\title{Some applications of the Lambert W function \\
to classical statistical mechanics.}
\author{Jean-Michel Caillol}
\affiliation{Laboratoire de Physique Th\'eorique \\
UMR 8267, B\^at. 210 \\
Universit\'e de Paris-Sud \\
91405 Orsay Cedex, France}
\email{Jean-Michel.Caillol@th.u-psud.fr}
\date{\today}
\begin{abstract}
We apply the recently defined Lambert  W function to some problems of
classical statistical mechanics, i.e. the Tonks gas
and a fluid of classical particles interacting via repulsive pair 
potentials. The latter case is considered both from the  point of view
of the standard theory of liquids and in the framework of a  field theoretical
description.
Some new mathematical properties of the  Lambert  W function are
obtained by passing. 
\end{abstract}
\maketitle
\section{Introduction}
\label{intro}

The Lambert W function is defined as the multivalued inverse of the function $w
\mapsto we^w$. Its mathematical properties have been explored only quite 
recently
after its implementation in the mathematical library of the computer algebra 
program \textit{Maple}.
The history, applications and properties of Lambert $W$ are reviewed in Corless
\textit{et al}. \cite{Corless1} Mathematical developments and applications to
physics (mainly in quantum mechanics and electrostatics) can be found in
refs.~\onlinecite{Corless1,Corless2,Corless3,Corless4}. Since  W is a
very simple function, all its applications to physics or other fields 
are not exhausted, and, after a short enquiry, I have discovered that many of my
colleagues, aware or unaware of its name, have met the function in their own
works.\cite{Viot}
 
Here we reformulate some old and new problems of
the statistical mechanics of classical liquids in terms of Lambert W.

Our paper is organized as follows. In  sec.\ II we  commence by
refreshing the reader ideas with some known mathematical properties of 
$W(z)$. We focus on the principal branch  $W_0(z)$ of the Lambert
function and the related function $U_0(z)=W_0(z)+W_0^2(z)/2$, both being 
of prime
importance for their applications to the problems of statistical mechanics
considered in this article.
Some new mathematical properties of  $W_0(z)$ and $U_0(z)$ used in the paper
are presented in the appendix (the Legendre transforms of the real functions
$x \in \R \; , x \mapsto W_0(\exp(x))$ and $x \mapsto  U_0(\exp(x))$
are computed  and the dispersion relations for  $W_0(z)$ and $U_0(z)$
established).

In sec.\ III we reformulate the properties of the Tonks gas,\cite{Tonks}
i.e. a
one-dimensional (1D)  classical fluid of hard rods, in terms of the principal
branch $W_0(z)$ of the Lambert function.
Our presentation is compared with the seminal work of Hauge
and Hemmer on this system.\cite{Hauge} The introduction of Lambert $W_0(z)$ 
allows to recover all the properties of the model in a unified treatment.

In sec.\ IV we focus on the second system considered By Hauge and Hemmer in 
ref.~\onlinecite{Hauge}, i.e. a classical $3D$ fluid 
made of particles with 
intermolecular pair repulsions of the form
\begin{equation}
\label{weak}
\varphi(r)=\gamma^3 \phi (\gamma r)
\end{equation}
where $\phi$ is a positive integrable function. 
The extension to arbitrary dimension
is trivial. In the
limit $\gamma \to 0$ which will be considered,
we deal with infinitely weak and infinitely long range repulsive interactions.
Once again,
the use of the functions $W_0(z)$ and $U_0(z)$ is of a great help for
describing in a unified manner all the known properties of this model.

Sec.\ V contains  new material. We consider the field theoretical
representation of a fluid of particles interacting via repulsive potentials 
in the framework of the formalism developed in ref.~\onlinecite{Caillol}. 
It is shown that at the mean field (MF) level of the theory the expressions 
for the pressure and the density
are identical to those  derived in sec.\ IV for the fluid with
weak long range repulsive interactions. In other words, the
MF approximation is exact in the case of infinitely 
long range interactions, a satisfactory, although expected result.
 We conclude in Sec.\ VI.

\section{A digest on Lambert W}
\label{Prole}
In this section we summarize the main mathematical properties of
the Lambert $W(z)$ function.  $W(z)$ is defined as the root of
\begin{equation}
\label{W}
 W(z) e^{ W(z)}=z \; ,
\end{equation} 
where $z$ is a complex number. $W(z)$ is a multivalued function which has
been studied recently in Refs.~\onlinecite{Corless1,Corless2,
Corless3,Corless4}. The different branches of Lambert W 
(i.e.\ the different possible
solutions of eq.\ (\ref{W})) are labeled by an integer 
$k=0,\pm 1,\pm 2,$ etc. When $z$ is a real number eq.\ (\ref{W}) can have
either two real solutions for  $0 > z >-\exp(-1)$,
 in which case they are
$W_0(z)$ and $W_{-1}(z)$, or it can have only one real solution for 
$ z \geq 0$, this being $W_0(z)$ while $W_{-1}(z)$ is now complex, or no real
solutions for $- \infty < z < -\exp(-1)$. 
$W_0(z)$ and $W_{-1}(z)$ are the only branches of $W$  
that take on real values.

A brief survey of the properties of the principal branch $W_0(z)$ will be
sufficient for our purpose.
Firstly, $W_0(z)$ is analytic at $z=0$. This follows from the Lagrange 
inversion theorem\cite{Cara}. Its power series about the point $z=0$
reads as
\begin{equation}
\label{Wseries}
W_0(z) = \sum_{n=1}^{\infty} \frac{(-n)^{n-1}}{n!} z^n
\; .
\end{equation}
The radius of convergence of the series\ (\ref{Wseries}) is equal to $e^{-1}$ as
it is easily shown using the Cauchy test and,
within the circle of convergence,   $\vert W(z) \vert <1$.
As a consequence of
the relation\ (\ref{W}) one  has
\begin{subequations}
\label{Wprime}
\begin{eqnarray}
\label{Wprimea}
W_0'(z)&=& \frac{W_0(z)}{z(1+W_0(z))} \; , \\
\label{Wprimeb}
&=& \sum_{n=1}^{\infty} \frac{(-n)^{n-1}}{(n-1)!} z^{n-1} \; ,
\end{eqnarray}
\end{subequations}
where the radius of convergence of the series\ (\ref{Wprimeb}) 
is again equal to $e^{-1}$. Lagrange theorem gives more than 
eq.\ (\ref{Wseries}); 
it allows
to obtain power series for $W_0^{\alpha}(z)$, $1/(1+W_0(z))$, $\exp(\alpha
W_0(z))$, etc for $z$ within the circle of convergence of $W_0(z)$. 
\cite{Corless3}
In particular, the power series of the function $U_0(z)$ defined to be 
$W_0(z) + W_0(z)^2/2$ is given by
\begin{equation}
\label{Useries}  
U_0(z)= \sum_{n=1}^{\infty} \frac{(-1)^{n-1}n^{n-2}}{n!} z^n 
\; \; (\vert z \vert < e^{-1}) \; .
\end{equation}
Note that if follows from eq.\ (\ref{Wprimea}) that
\begin{equation}
\label{Uprime}
zU_0'(z) = W_0(z) \; .
\end{equation}

$W_0(z)$ has a second-order branch point at $z=-e^{-1}$ which it shares with 
both
$W_1(z)$ and $W_{-1}(z)$ and its branch cut is conveniently chosen to be 
$\{ z : -\infty < z \leq -e^{-1} \}$ with the convention that $W_0(z)$ is 
defined
on the upper lip of the branch cut.
The behavior of $W_0(z)$ about the
branch point is given by the series\cite{Corless1}
\begin{equation}
\label{WseriesB}
W_0(z) = -1 + p -\frac{1}{3}p^2
+ \frac{11}{72}p^3 + \ldots
\; ,
\end{equation}
where $p=\sqrt{(2(ez+1)}$ (the series converges for 
$\vert p \vert <\sqrt{2}$).

In physics, it is quite common (and sometimes enlighting)  to write 
dispersion relations 
for functions having a branch cut such that $W_0(z)$ and $U_0(z)$.
\cite{Dennery} We show in the appendix that, in the present case,
these relations can be recast under the form
\begin{subequations}
\label{dispersion}
\begin{eqnarray}
\label{dispersionW}
W(U)_0(z)&=& \int_{-\infty}^{-e^{-1}} \; 
g_{W(U)}(s) \ln (1-\frac{z}{s}) \; ds
 \; , \\
\label{dispersionU}
g_{W(U)}(s)&=& - \frac{1}{\pi} \frac{d}{ds}\Im [W(U)(s)]\; ,
\end{eqnarray}
\end{subequations}
where $z$ is an arbitrary complex number which however is \textit{not}
on the branch cut. 
These equations convey the interpretation of $W_0(z)$ (resp. $U_0(z)$)
as the two-dimensional complex electric potential created by a distribution 
of charges
$g_{W}(s)$ (resp. $g_{U}(s)$) located on the branch cut. Moreover, we
show in the appendix that the distribution $g_{W}(s)$ is normalized to unity
(cf eq.\ (\ref{normaW})) while the total charge  of the charge distribution
$g_{U}(s)$ is infinite.
Other physical interpretations of the relations\ (\ref{dispersion}) will be
discussed in forthcoming developments.

\section{The Tonks gas}
\label{Tonks}
In this section we consider the Tonks gas, i.e. a 1D fluid of hard rods of
length $\sigma$.\cite{Tonks} The equation of state (EOS) is known exactly:
\begin{equation}
\chi \equiv \beta P = \frac{\rho}{1-\rho \sigma} \; .
\end{equation}
($P$   pressure, $\beta=1/kT$, $T$ temperature, $\rho$ number density). Of
course $0<\rho \sigma <1$. For $\rho \sigma \to 0$ one recovers the EOS
 of the ideal
gas, while for $\rho \sigma \to 1$, which corresponds to the close packing of the
hard rods, the pressure diverges. Less trivial is the expression of the 
pressure as a function of the activity $z$. Recall that $z=\exp(\beta \mu)$
where $\mu$ is the chemical potential and we have
assumed that the deBroglie thermal wavelength $\Lambda=1$. Starting from the
thermodynamic relation
\begin{equation}
\label{thermo}
\rho=z\frac{d}{dz}\chi \; ,
\end{equation}
it is not difficult to obtain the relation\cite{Hauge}
\begin{equation}
\label{relaTonks}
\sigma \chi e^{\sigma \chi}=\sigma z \; .
\end{equation}
\textit{A priori} Eq.\ (\ref{relaTonks}) is  valid only 
for real positive values of $z$ but 
it allows to define the pressure $\chi(z)$ in the
complex plane of the activities by analytic continuation.
The authors of ref.~\onlinecite{Hauge} did not know the Lambert function as we
do; clearly one has
\begin{equation}
\chi(z)=\frac{1}{\sigma} W_0(\sigma z) \; ,
\end{equation}
since the other branches $ W_k$ of Lambert W will not give a real 
pressure
for real chemical potentials $\mu$. 
The Mayer expansions for the pressure and
the density of the Tonks gas are therefore
obtained by copying out eqs.\ (\ref{Wseries}) and\
(\ref{Wprimeb}):
\begin{subequations}
\label{MayerT}
\begin{eqnarray}
\label{MayerT1}
 \chi(z)&=& \sum_{n=1}^{\infty} b_n \; z^n\; , \\
\label{MayerT2}
\rho(z) &=& \frac{1}{\sigma} \frac{W_0(\sigma z)}{1+ W_0(\sigma z)}
=   \sum_{n=1}^{\infty} n b_n \; z^n \; , \\
\label{MayerT3}
b_n&=&  \frac{ (-n \sigma)^{n-1}}{n!}        \; .
\end{eqnarray}
\end{subequations}

The results of sec.\ II on Lambert W enable us to conclude that
\begin{enumerate}
\item The radius of convergence of the Mayer series\ (\ref{MayerT1}) and\
(\ref{MayerT2}) is $R=1/e\sigma$. Within the circle of convergence $\vert
\chi(z) \vert <\sigma^{-1}$.
\item The pressure $\chi(z)$ is singular at the point $z=-R$ of the circle of
convergence, i.e. the branch point
of $W_0( \sigma z)$, in agreement with one of 
the conclusions of the second theorem of Groeneveld.\cite{Groeneveld} 
\item $\chi(z)$ has a branch cut on the negative real axis:
$-\infty<s\leq -R$.
 It can be identified with the distribution of zeros of the grand partition
function in the thermodynamic limit.\cite{Yang,Lee} The
distribution of the Yang-Lee zeros is given by $g(s)=g_W(\sigma s)=
-\Im W_0'(\sigma s)/\pi$
(cf eqs.\ (\ref{dispersionW}) and\ (\ref{DispersionW})).
$g(s)$ is a positive and
increasing function over the interval $-\infty <s < -R$ which behaves as
$g(s)\sim -1/(\sigma s \ln\vert s
\vert)^2)$ for $s \to -\infty$ and as $g(s) \sim \sqrt{e/2}/ (\pi x^{1/2})$
for $x=-R-x$, $x \to 0+$; therefore the integral of $g(s)$ 
over the interval $-\infty <s < -R$ is convergent and equal to $\sigma$
(cf eq.\ (\ref{normaW})).
\item For an arbitrary $z$ (however \textit{not} on the branch cut) the pressure
$\chi(z)$ may be seen as the $2D$ complex electrostatic potential created by the
charge distribution $g(s)$. It follows from 
eqs.\ (\ref{dispersionW}) and\ (\ref{DispersionW}) that $\chi(z)$ takes the
form proposed by Yang and Lee\cite{Yang,Lee}
\begin{equation}
\label{DispersTonks}
\chi(z)= \int_{-\infty}^{-R}  \; 
g(s) \ln (1-\frac{z}{s}) \; ds 
 \; \; (\forall z \; \text{ not on the cut } -\infty <s \leq -R \; )
\end{equation}
A close examination of their paper
reveals that the derivation of eq.\ (\ref{DispersTonks})
given by Hauge and Hemmer is restricted to a point $z$ inside the circle
of convergence.
\end{enumerate}
It is amusing to check the various conclusions of  second theorem of 
Groeneveld.\cite{Groeneveld} Recall that for a d-dimensional classical fluid 
of particles interacting via
positive pair potentials $\varphi(r_{ij})\geq 0 $ such that (twice)
the second virial coefficient
\begin{equation}
\label{virial}
f=\int d^d \vec{r} \; (\exp(-\beta \varphi(r))-1) \; ,
\end{equation}
converges, the radius of convergence $R$ of the virial series\ (\ref{MayerT1})
and the Mayer coefficients $b_n$ satisfy to the following inequalities

\begin{subequations}
\label{Groe}
\begin{eqnarray}
\label{Groe1}
 \frac{1}{e\vert f \vert} &\leq & R \leq \frac{1}{\vert f \vert}    \; , \\
\label{Groe2}
 \frac{1}{n}&\leq & \frac{b_n}{f^{n-1}} \leq \frac{n^{n-2}}{n!}    \; .
\end{eqnarray}
\end{subequations}
In the present case $f=2b_2=-\sigma$, and one easily checks that 
$R=1/e\sigma$ and $b_n$ as given by eq.\ (\ref{MayerT3}), do satisfy
to the Groeneveld inequalities\ (\ref{Groe}).

We end this section by determining the Helmoltz free energy per unit volume 
$\beta f(\rho)$ as the
Legendre transform of $\beta P$ viewed as a function of 
the reduced chemical potential
$\nu \equiv \beta \mu=\ln z$ (see eg ref.~\onlinecite{Caillol2}).
It follows from the eq.\ (\ref{Legendref}) of the appendix that
\begin{eqnarray}
\beta f(\rho)&=&\sup_{\nu \in \R }( \nu  \rho-\chi(e^{\nu})) \; \nonumber \\
\label{po}
&=& \rho(\ln\rho\Lambda -1) -\rho \ln(1-\rho\sigma) \; .
\end{eqnarray}
Note that we have restored $\Lambda$ to make the argument of the 
$\ln$ dimensionless.
$\beta f(\rho)$ is a strictly convex function of the
density defined for $0< \rho\sigma <1$. The first term in the RHS 
of the equation is the
Helmoltz free energy per unit volume of the ideal gas while the second one
is the excess free energy. The analytic continuation of $\beta f(\rho)$ to
complex densities is obtained by defining $\beta f(\rho)$ according to
eq.\ (\ref{po}) where $\ln$ is the principal branch of the natural logarithm 
for $\beta f(\rho)$ must take on real values for
 $\rho \in \R$, $0<\rho \sigma <1$.
$\beta f(\rho)$ is then a multivalued function with  two branch cuts
 $-\infty <\rho \leq 0$
 and  $\sigma^{-1}\leq \rho <+\infty$.

\section{Weak and long range repulsion}
\label{Weak}
The EOS of a gas of particles interacting via the pair potential\ (\ref{weak})
is known exactly and given by \cite{Hauge}

\begin{equation}
\label{chiw}
\chi \equiv \beta P = \rho + \frac{a}{2} \rho^2\; ,
\end{equation}
where
\begin{equation}
a= \beta \widetilde{\phi}(0) \equiv
\int d^3 \vec{r} \; \beta \phi(r) 
\end{equation}
is a positive, $\gamma$-independent constant. Here $\rho$ can take on all non
negative real values. 
It is not difficult to obtain the relation between the density
and the activity which reads as \cite{Hauge}

\begin{equation}
z=\rho e^{a \rho}.
\end{equation}
Therefore, since for a real $z$ the density must be real
\begin{equation}
\label{lamb1}
\rho= \frac{1}{a} W_0(a z) \; ,
\end{equation}
and, by integration of the thermodynamic relation\ (\ref{thermo})
\begin{equation}
\label{lamb2}
\chi(z)=\frac{1}{a} U_0(a z) \; .
\end{equation}
The pressure and the density for complex activities
are obtained by an analytical continuations of 
eqs.\ (\ref{lamb1}) and\ (\ref{lamb2}). 
The virial series for $\chi(z)$ and $\rho(z)$ follow from eqs.\ (\ref{Wseries})
and\ (\ref{Useries})
\begin{subequations}
\begin{eqnarray}
\label{toto}
 \chi(z)&=&  
 \sum_{n=1}^{\infty} \frac{(-a)^{n-1}n^{n-2}}{n!} z^n
 \; , \\
\rho(z) &=&  
\sum_{n=1}^{\infty} \frac{(-a)^{n-1}n^{n-1}}{n!} z^n
\; .
\end{eqnarray}
\end{subequations}
The radius of convergence of these two series is $R=(ae)^{-1}$ and it satisfies
to the first Groeneveld inequality\ (\ref{Groe1}). One also checks that the 
coefficients of the power series\ (\ref{toto}) do satisfy to the second
Groeneveld inequality\ (\ref{Groe2}). $\chi(z)$ has a branch cut  on 
the negative real axis: $-\infty<s\leq -R$. For any complex activity $z$
\textit{not} on the cut, it can be written once again
under the form proposed by Yang and
Lee \cite{Yang,Lee} (cf eq.\ (\ref{DispersionU}))
\begin{equation}
\label{Dispersweak}
\chi(z)= \int_{-\infty}^{-R}  \; 
g(s) \ln (1-\frac{z}{s}) \; ds 
 \; ,
\end{equation}
where
\begin{equation}
\label{gweak}
g(s) = g_U(as)=-\frac{1}{\pi a s} \Im W_0(as) \; .
\end{equation}
It follows from the analysis given in  the appendix that the function $g(s)$ 
is not integrable on the cut (it behaves as $-1/as$ for $s \to -\infty$).
Recall that, strictly speaking, the
theory of Yang and Lee is valid only for interactions with a hard core
contribution. In this case the grand partition function $\Xi$
 for a finite volume $V$ 
is a polynomial in $z$. The distribution $g(s)$ of its zeros  
after the passage to the thermodynamic limit allows to rewrite $\chi(z)\equiv
\ln \Xi/V$ under
the form\ (\ref{Dispersweak}), where $g(s)$ is normalizable. 
\cite{Hauge,Yang,Lee}
For soft interactions as those considered in
this section,
the reasoning breaks down, leaving us with a non normalizable distribution
$g(s)$ (one can pack an infinite number of particles in a finite volume and the
density $ \rho $ is not bounded).

Finally the Helmoltz free energy per unit volume 
$\beta f(\rho)$ is computed as the
Legendre transform of $\chi(e^{\nu})$. 
It follows from the eq.\ (\ref{Legendreh}) of the appendix that
\begin{eqnarray}
\beta f(\rho)&=&\sup_{\nu \in \R }( \nu  \rho-\frac{1}{a}
U_0(a e^{\nu})) \; \nonumber \\
\label{poweak}
&=& \rho(\ln\rho\Lambda^3 -1) + \frac{a}{2}\rho^2 \; .
\end{eqnarray}
$\beta f(\rho)$ is a strictly convex function of $\rho$ on the interval
$0<\rho <   + \infty$.
 
\section{Repulsive interactions: a field theoretical approach}
\label{KSSHE} 

We consider the statistical mechanics of a system made of $N$ classical point
particles interacting via pair potentials of the form
\begin{equation}
v(r)=\varphi_0(r) + \varphi(r) \; ,
\end{equation}
where $\varphi_0(r)$ is some reference potential (eg a hard sphere repulsion
for instance)
and $\varphi(r)$
admits a positive, well
defined  Fourier transform $\widetilde{\varphi}(k) \geq 0$.
$\varphi_0(r)$ and  $\varphi(r)$ are supposed to meet all the requirements
which are necessary for the existence of a thermodynamic limit (TL).
\cite{Ruelle}

We denote by $\Omega$ the domain 
occupied by the molecules of the fluid.
It will be convenient to assume that $\Omega$ is a cube of side $L$ with
periodic boundary conditions (PBC). In a given configuration
$\omega=(N;\vec{r}_1 \ldots \vec{r}_N)$
the microscopic density of particles reads as 
\begin{equation}
\label{rho}
\widehat{\rho}(\vec{r})=
\sum_{i=1}^{N} \delta^{(3)}(\vec{r}-\vec{r}_i) \; ,
\end{equation}
and its Fourier transform
\begin{equation}
\label{rhok}
\widehat{\rho}_{\vec{k}}=
\sum_{i=1}^{N} \exp(- i \vec{k}.\vec{r}_i) \; .
\end{equation}
The configurational potential energy of the system can be decomposed as
\begin{equation}
\label{V}
\beta V(\omega)=\beta V_0(\omega) -N\nu_S + \frac{\beta}{2}
\langle \widehat{\rho} \vert \varphi \vert \widehat{\rho} \rangle\; ,
\end{equation}
where $V_0(\omega)$ is the configurational energy of the reference system,
$\nu_S \equiv \beta \varphi(0)$ a self-energy contribution and
 
\begin{equation}
\langle \widehat{\rho} \vert \varphi \vert \widehat{\rho} \rangle \equiv
\int_{\Omega} d^{3} \vec{r}_1 d^{3} \vec{r}_2 \; 
\widehat{\rho}(\vec{r_1})
 \varphi(\vec{r}_1, \vec{r}_2)  \widehat{\rho}(\vec{r_2}) \; .
\end{equation} 
We shall 
work in the grand canonical (GC) ensemble. We denote by $\mu$ 
the chemical potential and by $\psi(\vec{r})$ an external potential. The local
chemical potential will be noted $\nu(\vec{r})=\beta(\mu-\psi(\vec{r}))$.
Performing a Kac-Siegert-Stratonovich-Hubbard-Edwards (KSSHE) transform 
\cite{Kac,Siegert,Strato,Hubbard1,Edwards}
one can show that the GC partition function can be recast under the
form of a functional integral\cite{Caillol}
\begin{equation}
\label{Xi}
\Xi[\nu]= {\cal N}^{-1} \int {\cal D} \xi
 \exp(-{\cal H}[\xi])
\: ,
\end{equation}
where ${\cal D} \xi$ is the functional measure and the action
${\cal H}[\xi]$ of the KSSHE field theory reads as

\begin{equation}
\label{H}
{\cal H}[\xi]  = \frac{1}{2\beta} \left< 
\xi
 \vert \varphi^{-1}
\vert \xi \right>  - 
\log \Xi_{0}\left[\overline{\nu} + i \xi \right] \; ,
\end{equation}
where $\overline{\nu}=\nu + \nu_S$ and $\Xi_{0}$ is the GC partition function
of the reference system. The field $\xi$ which enter eqs.\ (\ref{Xi}) and\
(\ref{H}) is a real scalar field. Note that, in the general case where
 the sign of
$ \widetilde{\varphi}(k)$ is abitrary,
the action ${\cal H}$ involves  \textit{two}
real scalar fields $\xi_+$ and $\xi_-$ (or a complex field $\xi$).
\cite{Caillol}  Finally,
the normalization constant $ {\cal N}$ is given by
\begin{equation}
\label{N}
{\cal N}=  \int {\cal D} \xi \exp(-\frac{1}{2\beta} \left< 
\xi
 \vert \varphi^{-1}
\vert \xi \right> ) \; .
\end{equation}
In ref.~\onlinecite{Caillol} the potential $\varphi_0(r)$ of the reference
system was chosen to be a hard core potential of diameter $\sigma$. Here we
specialize to the case $\sigma \to 0 $ (or equivalently $\varphi_0 \equiv 0 $), 
i.e. we take for the reference system 
the ideal gas. All the conclusions of  ref.~\onlinecite{Caillol} remain
valid provided that the TL of our system is well behaved.
Note that the configuration energy\ (\ref{V}) can be rewritten as
\begin{equation}
V(\omega)=\frac{1}{L^3} \sum_{\vec{k}} \vert \widehat{\rho}_{\vec{k}} \vert^2
\widetilde{\varphi}(k) - \frac{N}{2}\varphi(0) \; .
\end{equation}
Therefore if $\widetilde{\varphi}(k) \geq 0 \text{ for all } k$ and
$ \varphi(0)>0$ (which will be assumed henceforth) then $V(\omega)\geq -NB$ with
$B (\equiv \varphi(0)/2) >0$, i.e.
the system is H-stable in the sense of Ruelle and the TL
exists.\cite{Ruelle} Note that, conversely, if $\widetilde{\varphi}(k=0) <0$,
the system does not have a thermodynamic behavior and the introduction of
a repulsive hard core is mandatory to ensure the existence of a TL.
Such potentials (i.e. hard core plus an
attractive tail) can yield a liquid-vapor
transition as explained in ref.~\onlinecite{Caillol}; in the case considered
here (soft repulsive tail), the possibility of such a transition has to be
ruled out.

With the choice $\varphi_0 \equiv 0$ the KSSHE action reads now
\begin{equation}
\label{Hbis}
{\cal H}[\xi]  = \frac{1}{2\beta} \left< 
\xi
 \vert \varphi^{-1}
\vert \xi \right>  - 
\int_{\Omega}d^3\vec{r} \; e^{\overline{\nu}(\vec{r}) + i \xi(\vec{r})}\; .
\end{equation}
We turn now our attention to the mean field (MF) level of the theory.
The MF or saddle point approximation is defined by the equation
\begin{equation}
\label{xiMF}
\Xi_{MF}(\nu)\equiv \exp(-{\cal H}(\overline{\xi})) \; ,
\end{equation}
where at $\xi=\overline{\xi}$ the action ${\cal H}$ is minimum. 
If there
are several local minima then one retains the absolute minimum.
The stationary condition 
\begin{equation}
    \label{statio}
\left.\frac{\delta {\cal H}}{\delta \xi(\vec{r})}
\right|_{\overline{\xi}}=0  
\end{equation}
can be recast under the form of the implicit integral equation
\begin{equation}
    \label{statio1}
    \overline{\xi}(\vec{r})=i \beta
    \int_{\Omega}d^3 \vec{r}' \; \varphi(\| \vec{r}-\vec{r}' \|) \;
     e^{\overline{\nu}(\vec{r}') + i \overline{\xi}(\vec{r}')} \; .
\end{equation}
Moreover the MF density of the fluid is given by the density of 
the reference fluid -here the ideal gas- at
the local chemical potential $\overline{\nu}(\vec{r}) + i
\overline{\xi}(\vec{r})$\cite{Caillol}, i.e.

\begin{equation}
\label{rhoMF}
\rho_{MF}(\vec{r})=e^{\overline{\nu}(\vec{r}) + i \overline{\xi}(\vec{r})}
\; .
\end{equation}

In the case of a homogeneous system to which we stick from now,
(i.e. $\psi(\vec{r})\equiv 0$)
the solution of eq.\ (\ref{statio1}) is clearly
\begin{equation}
\label{xiMF1}
\overline{\xi}= i W_k(\lambda z ) \; ,
\end{equation}
where $W_k$ is some branch of Lambert W, $z$ the activity  and
\begin{equation}
\label{lambda}
\lambda=\beta \widetilde{\varphi}(0) e^{\nu_S} \; ,
\end{equation}
from which it follows that 
\begin{equation}
\label{rhoMF2}
\rho_{MF}=W_k(\lambda)/\beta \widetilde{\varphi}(0) \; .
\end{equation}
It remains to determine the branch of W. For real activities $z$, the MF density
should be a positive real number; it follows then from eq.\ (\ref{rhoMF2}) that,
necessarily, $k=0$.
The MF pressure is easily derived from eq.\ (\ref{xiMF}) with the result
\begin{eqnarray}
\label{chiMF}
\chi_{MF}(z)&=& \ln \Xi_{MF}/\Omega \; , \nonumber \\
&=&\frac{1}{\beta \widetilde{\varphi}(0)}
 U_0(\lambda z) \; .
\end{eqnarray} 
Note that the MF equation of state takes the familiar form 
$\chi_{MF}(\rho) =\rho + \beta \widetilde{\varphi}(0)\rho^2/2$. 
Finally, the MF free energy is obtained by a Legendre transform of $\chi(z)$,
which gives
 \begin{equation}
\label{fMF}
\beta f_{MF} = \rho(\ln(\rho \Lambda^3)-1) -\rho \nu_S +
 \beta \widetilde{\varphi}(0)\rho^2/2 \; ,
\end{equation}
in agreement with the general expression of ref.~\onlinecite{Caillol}.  

Some comments are in order.
\begin{enumerate}
\item 
The Taylor series
of $\chi_{MF}(z)$ and $\rho_{MF}(z)$ in terms of the activity $z$ 
are once again given by eqs.\
(\ref{Wseries}) and\ (\ref{Useries}) with a radius of convergence equal to
$R=(\lambda e)^{-1}$. For an arbitrary complex activity $z$, the pressure
 $\chi_{MF}(z)$ is
given by the dispersion relation\ (\ref{Dispersweak}).
\item 
We have shown elsewhere\cite{Caillol} that the one-loop order approximation
of the KSSHE field theory coincides with the random phase approximation (RPA)
of the theory of liquids.\cite{Hansen} As well known, for long range potentials of
the form\ (\ref{weak}) the RPA corrections to the pressure vanishes as
$\gamma \to 0$.\cite{Hansen} Note that, in this limit, $\nu_S \to 0$ and 
$\widetilde{\varphi}(0)=\widetilde{\phi}(0)$; thence, in the limit
$\gamma \to 0$, one recovers for the pressure, the free energy etc the expressions
derived in sec.\ IV. Stated  otherwise, the MF-KSSHE theory is exact for
infinitely weak and long range repulsive potentials.
\item We have shown in ref.~\onlinecite{Caillol} that the MF pressure and free
energy constitute \textit{exact} bounds. More precisely we have
\begin{subequations}
\begin{eqnarray}
 \forall z \in \R  \; \;   \chi(z)&\leq & \chi_{MF}(z) 
\equiv U_0(\lambda z)/\beta \widetilde{\varphi}(0)
\; , \\
\forall \rho >0  \; \;   \beta f(\rho) &\geq &  \beta f_{MF}(\rho) 
\equiv
\rho(\ln(\rho \Lambda^3)-1) -\rho \nu_S +
 \beta \widetilde{\varphi}(0)\rho^2/2
\; .
\end{eqnarray}
\end{subequations}
\end{enumerate}
\section{Conclusion}
In this paper we have discussed some applications of the Lambert W function to the
theory of liquids. In the case of $1D$ hard rods or infinitely weak long range
repulsive potentials, one obtains a close expression for  the complex pressure
$\chi(z)$ as a function of complex activities  in terms 
of either $W_0(z)$ or the related function $U_0(z)$.
The dispersion relations derived for $W_0(z)$ and
$U_0(z)$ in the appendix give a  rigorous justification to the heuristic 
formula proposed by Lee and Yang for $\chi(z)$.\cite{Lee,Yang}
We have also shown that, in the framework of the KSSHE field theory of liquids,
the MF pressure $\chi(z)$ of a gas of particles interacting via soft repulsive
potentials can be expressed in terms of the function $U_0(z)$.

\begin{acknowledgments}
The author likes to thank P. Viot for drawing his attention to
ref.~\onlinecite{Viot}.
\end{acknowledgments}
\appendix*
\label{appendix}
\section{New mathematical properties of Lambert W}
\subsection{Legendre transforms}
Let us first consider the real function $x \mapsto f(x)=W_0(e^x)$. By applying
twice the formula\ (\ref{Wprimea}) one finds
\begin{subequations}
\label{derif}
\begin{eqnarray}
\label{derif1}
 f'(x)&=&\frac{f(x)}{1 + f(x)}  \; , \\
\label{derif2}
f''(x)&=& \frac{f(x)}{(1 + f(x))^3}  \; .
\end{eqnarray}
\end{subequations}
It follows from eq.\ (\ref{derif2}) that $f''(x)>0$ for all $x \in \R$.
 Thence the
function $f$ is strictly convex and one can define its Legendre transform
${\cal L}_f(\rho)$ (see eg ref.~\onlinecite{Caillol2}). By definition
\begin{equation}
{\cal L}_f(\rho)=\sup_{x}(x \rho-f(x)),
\end{equation}
or, more precisely 
\begin{equation}
{\cal L}_f(\rho)=\overline{x} \rho  - f(\overline{x}) \; ,
\end{equation}
where $\overline{x}$ is (the unique) solution of
\begin{equation}
\label{l1}
\rho=f'(\overline{x})=\frac{f(\overline{x})}{1 + f(\overline{x})}  \; .
\end{equation}
Therefore ${\cal L}_f(\rho)$ is defined for $0 <\rho <1$. It is also a
strictly convex function, the Legendre transform of which is $f(x)$ (i.e. the
Legendre transform is involutive). In order to get the expression of 
${\cal L}_f(\rho)$ one notes that eq.\ (\ref{l1}) implies that
$f(\overline{x})=\rho/(1-\rho)$ and that $\overline{x}=f(\overline{x})+
\ln f(\overline{x})$ which yields 
 \begin{equation}
 \label{Legendref} 
 {\cal L}_f(\rho)=\rho \left(\ln \rho -1\right) -\rho \ln(\rho-1) \; .
\end{equation}
 
Similarly it can easily be shown that the function $x \mapsto h(x)=U_0(e^x)$ is
strictly convex and that its Legendre transform  ${\cal L}_h(\rho)$ is given by
 \begin{equation}
 \label{Legendreh} 
 {\cal L}_h(\rho)=\rho \left(\ln \rho -1\right) + \frac{\rho^2}{2} \; .
 \end{equation}
Note that ${\cal L}_h(\rho)$ is defined on $0< \rho <\infty)$. Over this
interval, ${\cal L}_h(\rho)$ is strictly convex.
\subsection{Dispersion relations}
\begin{figure}[!ht]
\begin{center}
\epsfig{figure=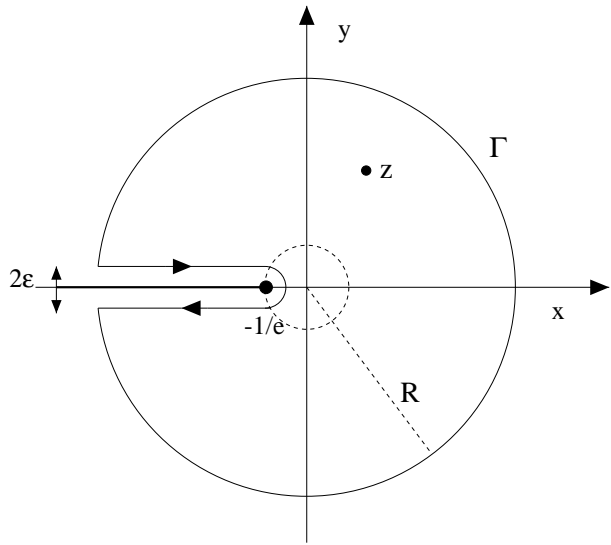,width=8cm,clip=}
\caption{\label{contour} Solid line: contour of integration of 
eq.\ (\ref{A1}).
The small circle (dashed line) is the circle of
convergence of $W_0(z)$ about $z=0$, the dot is the branch point and the thick
solid line  is the branch cut. The point $z$ is everywhere except on the cut. }
\end{center}
\end{figure}

We establish here  the dispersion relations for the functions
$W_0(z)$ and $U_0(z)$. Let us first consider $W_0(z)$.
It follows from the Cauchy theorem that 

\begin{equation}
\label{A1}
W_0(z)=\frac{1}{2\pi i} \int_{\cal{C}} \;
W_0(s) \{ \frac{1}{z-s} -\frac{1}{s} \} \; ds \; ,
\end{equation}
where $\cal{C}$ is the contour shown in fig.\ (\ref{contour}) and $z$
is  \textit{not} on the cut $(-\infty,-e^{-1})$. 
Indeed $W_0(0)=0$
as follows from eq.\ (\ref{Wseries}). We consider now eq.\ (\ref{A1}) in the
limit $\epsilon \to 0$ and $R \to \infty$.

The asymptotic formula for large (complex) $z$, i.e.
 \cite{Corless1,Corless4}
\begin{equation}
\label{asymp}
W_0(z) \sim \ln(z) - \ln(\ln(z))  \; ,
\end{equation} 
where $\ln z$ is the principal branch of the natural logarithm, 
ensures that the contribution to eq.\ (\ref{A1}) from the large circle
 $\Gamma$
tends to zero as its radius $R$ tends to infinity (Jordan lemma). Note that
 we
have substracted $W_0(0)=0$ from $W_0(z)$ in eq.\ (\ref{A1}) precisely
in order to obtain this property. Therefore we have

\begin{eqnarray}
\label{Cauchy}
W_0(z)=
\frac{1}{2\pi i} \int_{-\infty}^{-e^{-1}}  &\{& W_0(s+i \epsilon) \;
[\frac{1}{s+i \epsilon -z} -  \frac{1}{s+i \epsilon } ] \nonumber \\
&-& W_0(s- i \epsilon) \; [\frac{1}{s-i \epsilon -z} -  
\frac{1}{s-i \epsilon }
 ] \; \} \; ds\; .
\end{eqnarray}
As $\epsilon \to 0$, we can neglect the $\pm i\epsilon$ in the fractions that
enter the RHS of eq.\ (\ref{Cauchy})
(remember that $z$ is \textit{not} on the cut) and we get

\begin{equation}
\label{Cauchy2}
W_0(z)=\frac{1}{2\pi i} \int_{-\infty}^{-e^{-1}}
[W_0(s+i \epsilon) - W_0(s- i \epsilon) ] [
\frac{1}{s-z}- \frac{1}{s}] \; ds \; .
\end{equation}
With the convention that $W_0(s)$ is defined on the upper lip of the cut
and by
noting that $W_0(\overline{s})=\overline{W_0(s)}$ (for $s$ not on the cut)
 we infer from eq.\
(\ref{Cauchy2}) that
\begin{equation}
\label{Cauchy3}
W_0(z)=\frac{1}{\pi } \int_{-\infty}^{-e^{-1}}
\Im \left( W_0\left(s\right)\right) [
\frac{1}{s-z}- \frac{1}{s}] \; ds \; .
\end{equation}
The final step is to perform an integration by parts which yields
\begin{eqnarray}
\label{Dispers0}
W_0(z)&=& \big[ \Im \left( W_0\left(s\right)\right)\ln(1-\frac{z}{s})
\big]_{-\infty}^{-e^{-1}} \nonumber \\
&+&  \int_{-\infty}^{-e^{-1}} \frac{-1}{\pi} \frac{d}{ds}
 \Im \left( W_0\left(s\right)\right) \; 
 \ln(1-\frac{z}{s}) \; ds \; .
\end{eqnarray}
Since, in the one hand $W_0(-e^{-1})=-1$ and, on the other hand,
$\Im \left( W_0\left(-\infty\right)\right) =\pi$ as can
be obtained for instance from the asymptotic behavior\ (\ref{asymp}) of 
$W_0(z)$, then the first contribution to the RHS of eq.\ (\ref{Dispers0}) 
vanishes and we are left with
\begin{subequations}
\label{dispersW}
\begin{eqnarray}
\label{DispersionW}
W_0(z)&=& \int_{-\infty}^{-e^{-1}}  \; 
g_{W}(s) \ln (1-\frac{z}{s}) \; ds 
 \; , \\
\label{DispersionU}
g_{W}(s)&=& - \frac{1}{\pi} \frac{d}{ds}\Im [W_0(s)]\; .
\end{eqnarray}
\end{subequations}
Several comments are in place here.

(a) The function $g_{W}(s)$ is a positive increasing function on the cut
$(-\infty,-e^{-1})$.  For $s=-e^{-1} -x$ ($x\to +0)$,  
$g_{W}(s) \sim \sqrt{e/2}/ (\pi x^{1/2})$ as can
be
seen from the series expansion of  $W_0(z)$ about the branch point (cf
eq.\ (\ref{WseriesB})). In the other hand  $g_{W}(s) \sim -1/(s \ln\vert s
\vert)^2)$ for $s \to -\infty$ as can be infered from eq.\ (\ref{asymp}).
Thence the integral of $g_{W}(s)$ along the branch cut is convergent; more precisely
\begin{eqnarray}
\label{normaW}
\int_{-\infty}^{-e^{-1}} g_{W}(s) \; ds&=& \left[\Im \left(W_0\left(-\infty
\right) \right) - \Im \left(W_0\left(-e^{-1}
\right) \right) \right]/\pi \nonumber \\
&=& 1 \; .
\end{eqnarray}

(b) The dispersion relations for the function $U_0(z)$ are similar to eqs.\
(\ref{DispersionW}) for the reasoning presented above can be reproduced
without major changes. The distribution  $g_{U}(s)$ is again a positive function
on the interval $(-\infty, -e^{-1})$. It is given by

\begin{eqnarray}
\label{normaU}
g_{U}(s)&=&-\pi^{-1} d(\Im U_0(s))/ds \nonumber \\
&=&-\frac{\Im W_0(s)}{\pi s} \; ,
\end{eqnarray}
from which it follows that $g_{U}(-e^{-1})=0$ because $W_0(-e^{-1})=-1$.
 Moreover, for
$s \to -\infty$ we have $g_{U}(s)\sim -1/s$ with the consequence that
the integral of $g_{U}(s)$ along the branch cut is divergent to $+\infty$.


\end{document}